\documentstyle[aps,epsfig,balanced]{revtex}
\begin{document}

\title{Analytical and Numerical Studies of Noise-induced Synchronization
of Chaotic Systems}

\author{Ra\'{u}l Toral$^{1,2}$, Claudio R. Mirasso$^{2}$,
Emilio Hern\'{a}ndez-Garc\'{\i}a$^{1,2}$
and Oreste Piro$^{1,2}$}

\address{(1)Instituto Mediterr\'aneo de Estudios Avanzados,
IMEDEA\footnote{URL: http://www.imedea.uib.es/PhysDept\\e-mail:raul@imedea.uib.es}, CSIC-UIB.\\
(2) Departament de F\'{\i}sica, Universitat de les Illes Balears.
07071-Palma de Mallorca, Spain}

\date{April 5, 2001}

\maketitle

\begin{abstract}

We study the effect that the injection of a common source of noise
has on the trajectories of chaotic systems, addressing some
contradictory results present in the literature. We present
particular examples of 1-d maps and the Lorenz system, both in the
chaotic region, and give numerical evidence showing that the
addition of a common noise to different trajectories, which start
from different initial conditions, leads eventually to their
perfect synchronization. When synchronization occurs, the largest
Lyapunov exponent becomes negative. For a simple map we are able
to show this phenomenon analytically. Finally, we analyze the
structural stability of the phenomenon.

\end{abstract}

\hspace{2.0cm}

\begin{twocolumns}
{\bf The synchronization of chaotic systems has been the subject
of intensive research in the last years. Besides its fundamental
interest, the study of the synchronization of chaotic oscillators
has a potential application in the field of chaos communications.
The main idea resides in the hiding of a message within a chaotic
carrier generated by a suitable emitter. The encoded message can
be extracted if an appropriate receiver, one which synchronizes to
the emitter, is used. One of the conditions to be fulfilled in
order to achieve synchronization is that the receiver and the
emitter have very similar device parameters, hence making it very
difficult to intercept the encoded message. Although the usual way
of synchronizing two chaotic systems is by injecting part of the
emitted signal into the receiver, the possibility of
synchronization using a common random forcing has been also
suggested. However, there have been some contradictory results in
the literature on whether chaotic systems can indeed be
synchronized using such a common source of noise and the issue has
began to be clarified only very recently. In this paper we give
explicit examples of chaotic systems that become synchronized by
the addition of Gaussian white noise of zero mean. We also analyze
the structural stability of the phenomenon, namely, the robustness
of the synchronization against a small mismatch in the parameters
of the chaotic sender and receiver.}

\section{Introduction}
\label{introduction}

One of the most surprising results of the last decades in the
field of stochastic processes has been the discovering that
fluctuation terms (loosely called {\sl noise}) can actually induce
some degree of order in a large variety of non-linear systems. The
first example of such an effect is that of {\sl stochastic
resonance}\cite{BSV81,NN81} by which a bistable system responds
better to an external signal (not necessarily periodic) under the
presence of fluctuations, either in the intrinsic dynamics or in
the external input. This phenomenon has been shown to be relevant
for some physical and biological systems described by nonlinear
dynamical equations \cite{JSP70,GHJM98,LC98}. Other examples in
purely temporal dynamical systems include phenomena such as
noise-induced transitions\cite{HL84}, noise-induced
transport\cite{HB96}, coherence
resonance\cite{GDNH93,RS94,PK97,LS00}, etc. In extended systems,
noise is known to induce a large variety or ordering
effects\cite{OS99}, such as pattern formation\cite{GHS93,PBBR96},
phase transitions\cite{BPT,GPSB,MDZT,IGTS00}, phase
separation\cite{GLST98,IGTS99}, spatiotemporal stochastic
resonance\cite{JM95,MGB96}, noise-sustained
structures\cite{D89,SCSMW97}, doubly stochastic
resonance\cite{ZKS00}, amongst many others. All these examples
have in common that some sort of {\sl order} appears only in the
presence of the right amount of noise.

There has been also some recent interest on the interplay between
chaotic and random dynamics. Some counterintuitive effects such as
coherence resonance, or the appearance of a quasi--periodic
behavior, in a chaotic system in the presence of noise, have been
found recently\cite{PTMCG00}. The role of noise in standard
synchronization of chaotic systems has been considered in
\cite{A97,ADL00}, as well as the role of noise in synchronizing
non--chaotic systems\cite{J98,A99}. In this paper we address the
different issue of synchronization of chaotic systems by a common
random noise source, a topic that has attracted much attention
recently. The accepted result is that, for some chaotic systems,
the introduction of the same noise in independent copies of the
systems could lead (for large enough noise intensity) to a common
collapse onto the same trajectory, independently of the initial
condition assigned to each of the copies. This synchronization of
chaotic systems by the addition of random terms is a remarkable
and counterintuitive effect of noise and although some clarifying
papers have appeared recently, still some contradictory results
exist for the existence of this phenomenon of noise--induced
synchronization. It is the purpose of this  paper to give further
analytical and numerical evidence that chaotic systems can
synchronize under such circumstances and to analyze the structural
stability of the phenomenon. Moreover, the results presented here
clarify the issue, thus opening directions to obtain such a
synchronization in electronic circuits, for example for encryption
purposes. Common random noise codes have been used in spread
spectrum communication since a long time ago \cite{V95}. The main
idea is to mix a information data within a noisy code. At the
receiver, the information is recovered using a synchronized
replica of the noise code. More recently, the use of common noise
source has been also proposed as a useful technique to improve the
encryption of a key in a communication channel \cite{MA98}.

The issue of ordering effect of noise in chaotic systems was
considered already at the beginning of the 80's by Matsumoto and
Tsuda\cite{mt83} who concluded that the introduction of noise
could actually make a system less chaotic. Later, Yu, Ott and
Chen\cite{YOC90} studied the transition from chaos to non--chaos
induced by noise. Synchronization induced by noise was considered
by Fahy and Hamman\cite{FH92} who showed that particles in an
external potential, when driven by the same random forces, tend to
collapse onto the same trajectory, a behavior interpreted as a
transition from chaotic to non--chaotic behaviors. The same system
has been studied numerically and
analytically\cite{KV95,C96,KIM99}. Pikovsky \cite{Pik92} analyzed
the statistics of deviations from this noise-induced
synchronization. A paper that generated a lot of controversy was
that of Maritan and Banavar\cite{MB94}. These authors analyzed the
logistic map in the presence of noise:

\begin{equation}
x_{n+1}=4x_n(1-x_n)+\xi_{n}
\label{map}
\end{equation}
\\
where $\xi _{n}$ is the noise term, considered to be uniformly
distributed in a symmetric interval $[-W,+W]$. They showed that,
if $W$ was large enough (i.e. for a large noise intensity) two
different trajectories which started with different initial
conditions but used otherwise the same sequence of random numbers,
would eventually coincide into the same trajectory. The authors
showed a similar result for the Lorenz system (see section
\ref{lorenz}). This result was heavily criticized by
Pikovsky\cite{pik94} who proved that two systems can synchronize
only if the largest Lyapunov exponent is negative. He then argued
that the largest Lyapunov exponent of the logistic map in the
presence of noise is always positive and concluded that the
synchronization was, in fact, a numerical effect of lack of
precision of the calculation. The analysis of Pikovsky was
confirmed by Longa {\sl et al}. \cite{LCO96} who studied the
logistic map with arbitrary numerical precision. The criterion of
negative Lyapunov exponent has also been shown to hold for other
types of synchronization of chaotic systems and Zhou and
Lai\cite{ZL98} noticed that previous results by Shuai, Wong and
Cheng\cite{SWC97} showing synchronization with a positive Lyapunov
exponent were again an artifact of the limited precision of the
calculation.

In addition to the above criticisms, Herzel and Freund\cite{HF95}
and Malescio\cite{M96} pointed out that the noise used to simulate
Eq.(\ref{map}) and the Lorenz system in \cite{MB94} is not really
symmetric. While the noise in the Lorenz system is non--symmetric
by construction, in the case of the map, the non--zero mean arises
because the requirement $x_n \in (0,1),~\forall n$, actually leads
to discard the values of the random number $\xi_n$ which would
induce a violation of such condition. The average value of the
random numbers which have been accepted is different from zero,
hence producing an effective {\sl biased} noise, i.e. one which
does not have zero mean. The introduction of a non-zero mean noise
means that the authors of \cite{MB94} were altering essentially
the properties of the deterministic map. Furthermore, Gade and
Bassu\cite{GB96} argued that the synchronization observed by
Maritan and Banavar is due to the fact that the bias of the noise
leads the system to a non--chaotic fixed point. With only this
basis, they concluded that a zero--mean noise can never lead to
synchronization in the Lorenz system. The same conclusion was
reached by S\'anchez {\sl et al}.\cite{SMP97} who studied
experimentally a Chua circuit and concluded that synchronization
by noise only occurs if the noise does not have a zero mean. The
same conclusion is obtained in \cite{LP99} by studying numerically
a single and an array of Lorenz models, and in \cite{PL99} from
experiments in an array of Chua circuits with multiplicative
colored noise. Therefore, from these last works, a widespread
belief has emerged according to which it is not possible to
synchronize two chaotic systems by injecting the same noisy
unbiased, zero--mean, signal to  both of them.

Contrary to these last results (but in agreement with the
previously mentioned results
\cite{MA98,mt83,YOC90,FH92,KV95,C96,KIM99,Pik92}), Lai and
Zhou\cite{LZ98} have shown that some chaotic maps can indeed
become synchronized by additive zero--mean noise. A similar result
has been obtained by Loreto {\sl et al}. \cite{vulpi95}, and by
Minai and Anand\cite{MA98,min98,M00}, in the case where the noise
appears parametrically in the map. The implications to secure
digital communications have been considered in
\cite{MA98,min98,SM00}, and an application to ecological dynamics
in fluid flows is presented in \cite{Chaos2001}. An equivalent
result about the synchronization of Lorenz systems using a common
additive noise has been shown by the authors of the present paper
in\cite{TMHP99}. The actual mechanism that leads to
synchronization has been explained by Lai and Zhou\cite{LZ98}, see
also \cite{RWKK00}. As Pikovsky\cite{pik94} required,
synchronization can only be achieved if the Lyapunov exponent is
negative. The presence of noise allows the system to spend more
time in the ``convergence region" where the local Lyapunov
exponent is negative, hence yielding a global negative Lyapunov
exponent. This argument will be developed in more detail in
section \ref{maps}, where an explicit calculation in a simple map
will confirm the analysis. The results of Lai and Zhou have been
extended to the case of coupled map lattices\cite{BLT99} where
Pikovsky's criterion has been extended for spatially extended
systems.

In this paper we give further evidence that it is possible to
synchronize two chaotic systems by the addition of a common noise
which is Gaussianly distributed and not biased.  We analyze
specifically some 1-d maps and the Lorenz system, all in the
chaotic region.  The necessary criterion introduced in Ref.
\cite{pik94} and the general arguments of \cite{LZ98} are fully
confirmed and some heuristic arguments are given about the general
validity of our results.

The organization of the paper is as follows. In section \ref{maps} we present
numerical and analytical results for some 1-d maps, while section \ref{lorenz}
studies numerically the Lorenz system. In section \ref{structural} we analyze
the structural stability of the phenomenon, i.e. the dependence of the
synchronization time on the parameter mismatch. Finally, in section
\ref{conclusions} we
present the conclusions as well as some open questions relating the general
validity of our results.

\section{Results on maps}
\label{maps}
The first example is that of the map:
\begin{equation}
x_{n+1}= F(x_n) = f(x_n)+\epsilon \xi_n
\label{eq:2}
\end{equation}
where $\xi_n$ is a set of uncorrelated Gaussian variables of zero
mean and variance 1. As an example, we use explicitly
\begin{equation}
f(x)=\exp\left[-\left(\frac{x-0.5}{\omega}\right)^2\right]
\label{eq:3}
\end{equation}
Studying the convergence or divergence of trajectories of Eq.
(\ref{eq:2}) starting from different initial conditions under the
same noise $\xi_n$ is equivalent to analyzing the converge or
divergence of trajectories from two identical systems of the form
(\ref{eq:2}) driven by the same noise. We plot in Fig.(1) the
bifurcation diagram of this map in the noiseless case. We can see
the typical windows in which the system behaves chaotically. The
associated Lyapunov exponent, $\lambda$, is positive in these
regions. For instance, for $\omega=0.3$ (the case we will be
considering throughout the paper) it is $\lambda
\approx 0.53$. In Fig.(2) we observe that the Lyapunov exponent
becomes negative for most values of $\omega$ for large enough
noise level $\epsilon$. Again for $\omega=0.3$ and now for
$\epsilon=0.2$ it is $\lambda=-0.17$. A positive Lyapunov exponent
in the noiseless case implies that trajectories starting with
different initial conditions, but using the same sequence of
random numbers $\{\xi_n\}$, remain different for all the iteration
steps. In this case, the corresponding synchronization diagram
shows a spread distribution of points (see Fig.(3a)). However,
when moderate levels of noise ($\epsilon \gtrsim 0.2$) are used,
$\lambda$ becomes negative and trajectories starting with
different initial conditions, but using the same sequence of
random numbers, synchronize perfectly, see the synchronization
diagram in Fig.(3b). Obviously, the noise intensity in the cases
shown is not large enough such as to be able to neglect completely
the deterministic part of the map. Therefore, the synchronization
observed does not trivially appear as a consequence of both
variables becoming themselves identical to the noise term.

\begin{figure}[htb]
\epsfig{file=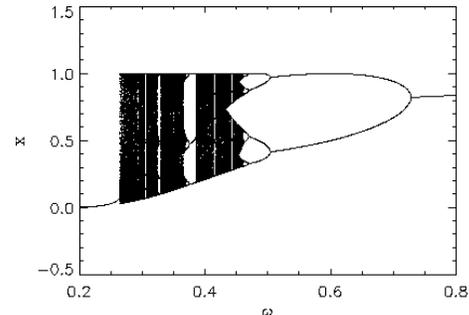,width=6.0cm}
\caption{
\label{fig1}
Bifurcation diagram of the map given by Eqs.(\ref{eq:2}) and
(\ref{eq:3}) in the absence of noise terms.}
\end{figure}

\begin{figure}[htb]
\epsfig{file=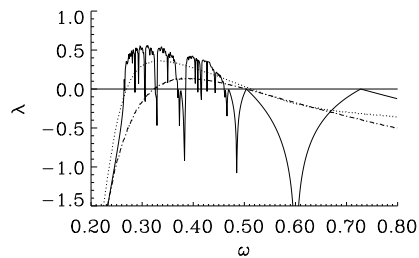,width=6.0cm}
\caption{
\label{fig2}
Lyapunov exponent
for the noiseless map ($\epsilon=0$, continuous line) and
the map with a noise intensity $\epsilon=0.1$ (dotted line) and
$\epsilon=0.2$ (dot-dashed line).}
\end{figure}

\begin{figure}[htb]
\epsfig{file=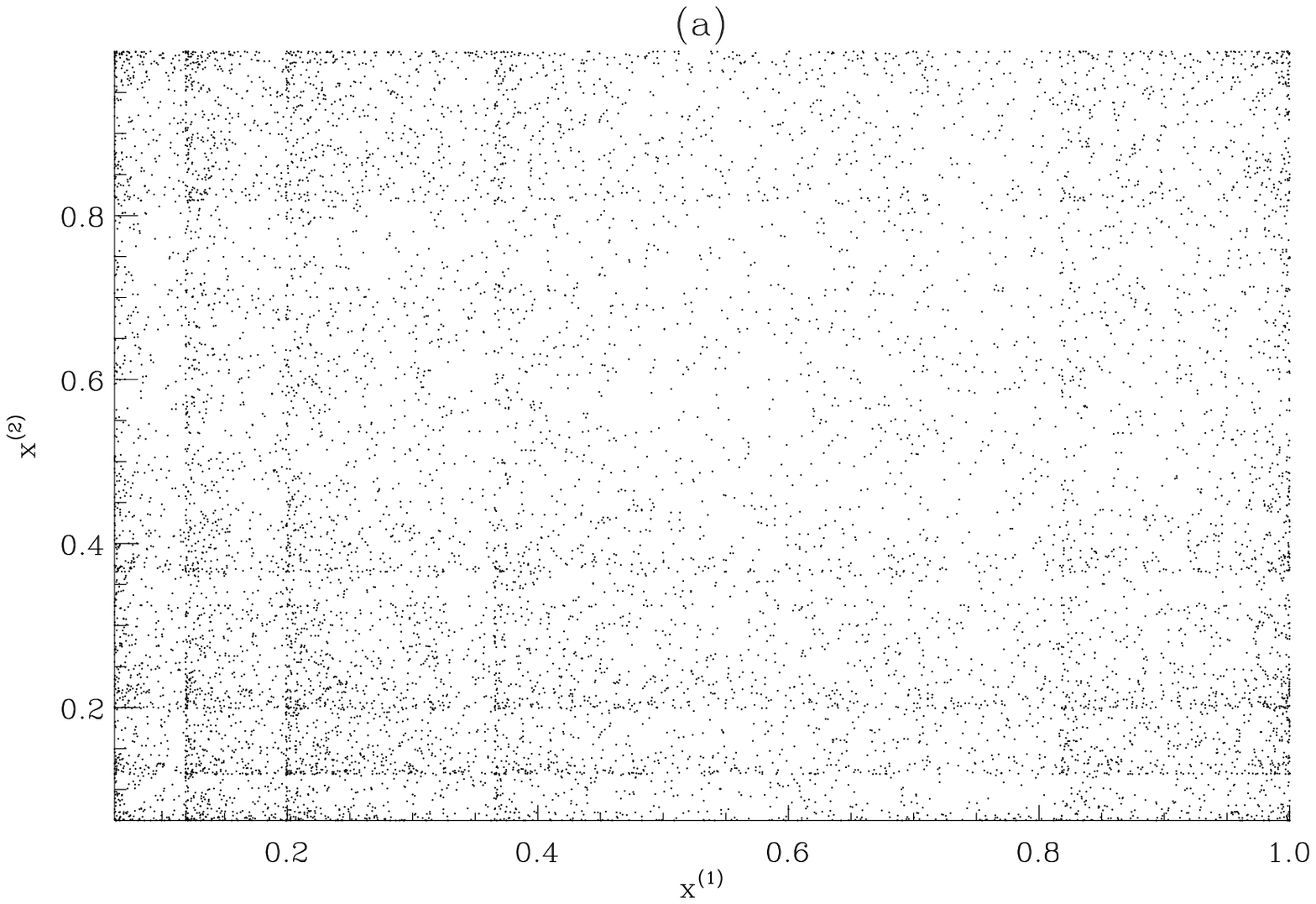,width=6.0cm}
\epsfig{file=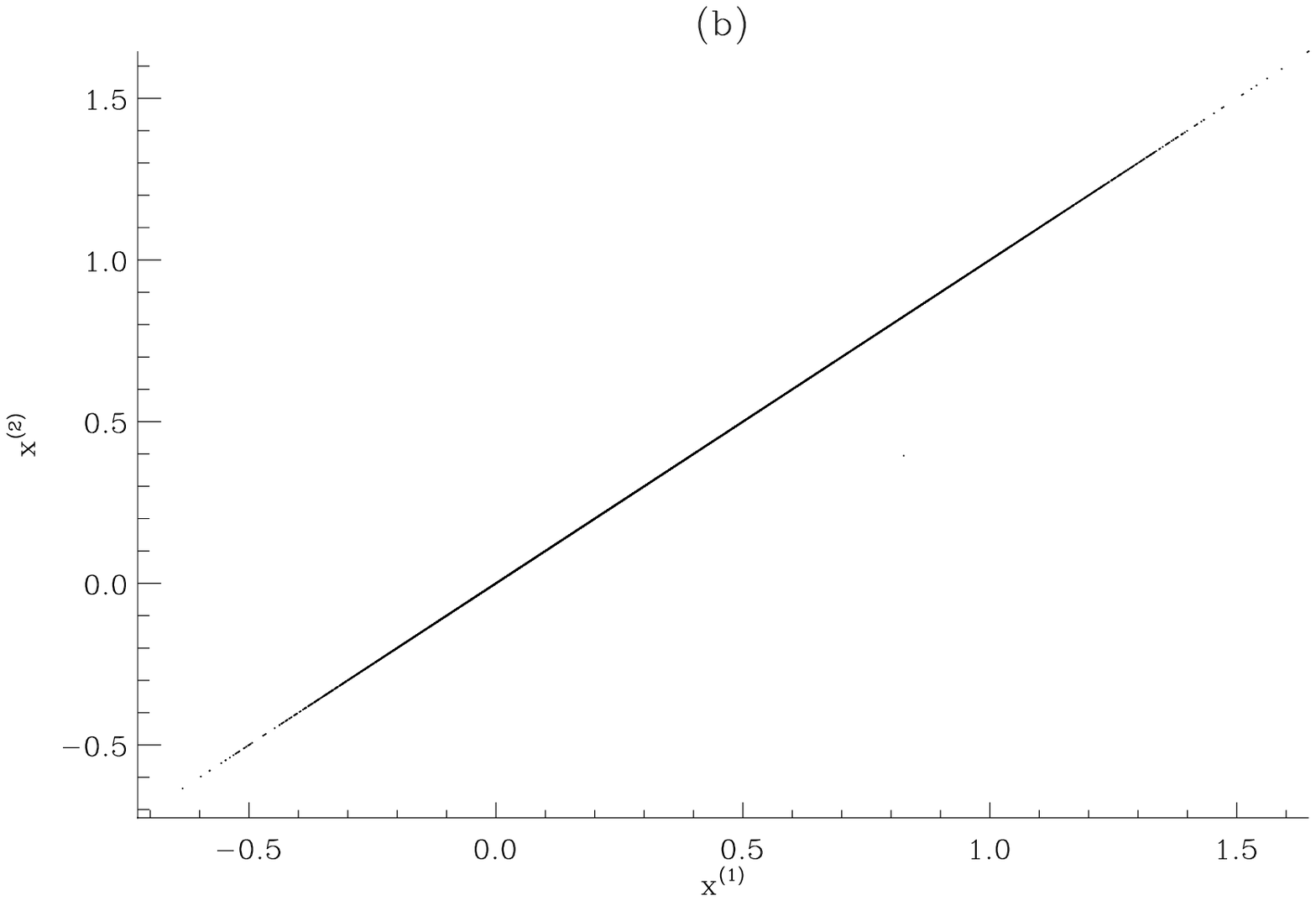,width=6.0cm}
\caption{
\label{fig3}
Plot of two realizations $x^{(1)}$, $x^{(2)}$ of of the map given
by Eqs. (\ref{eq:2}) and (\ref{eq:3}) with $\omega=0.3$. Each
realization consists of 10,000 points which have been obtained by
iteration of the map starting in each case from a different
initial condition (100,000 initial iterations have been discarded
and are not shown). In figure (a) there is no noise, $\epsilon=0$
and the trajectories are independent of each other. In figure (b)
we have used a level of noise $\epsilon=0.2$, producing a perfect
synchronization (after discarding some initial iterations). }
\end{figure}

According to \cite{pik94}, convergence of trajectories to the same
one, or lack of sensitivity to the initial condition, can be stated as
{\sl negativity of the Lyapunov exponent}. The Lyapunov exponent
of the map (\ref{eq:2}) is defined as

\begin{equation}
\label{lyapunov2}
\lambda =\lim_{N\rightarrow\infty }\frac{1}{N}\sum_{i=1}^N \ln|F'(x_i)|
\end{equation}
\\
It is the average of (the logarithm of the absolute value of) the
successive slopes $F'$ found by the trajectory.
Slopes in $[-1,1]$ contribute to $\lambda$ with
negative values, indicating trajectory convergence. Larger or
smaller slopes contribute with positive values, indicating trajectory
divergence.
Since the deterministic and noisy maps satisfy $F'=f'$  one
is tempted to conclude that the Lyapunov exponent is not modified by the
presence of noise. However, there is noise-dependence through the
trajectory values
${x_i}$, $i=1,...,N$. In the absence of noise, $\lambda$ is positive,
indicating trajectory separation.
When synchronization is observed, the Lyapunov exponent is negative,
as required by the argument in \cite{pik94}.

Notice that this definition of the Lyapunov exponent assumes a
fixed realization of the noise terms, and it is the relevant one
to study the synchronization phenomena addressed in this paper.
One could use alternative definitions \cite{vulpi95}. For
instance, if one considers the coupled system of both the $x$
variable and the noise generator producing $\xi$, then the largest
Lyapunov exponent of the composed system is indeed positive (and
very large for a good random number generator). This simply tells
us that there is a large sensitivity to the initial condition of
the composed system $(x,\xi)$ as shown by the fact that a change
of the seed of the random number generator completely changes the
sequence of values of both $\xi$ and $x$. We consider in this
paper the situation described by definition (\ref{lyapunov2}) with
fixed noise realization.

By using the
definition of the {\sl invariant measure on the attractor}, or {\sl
stationary probability distribution} $P_{st}(x)$, the Lyapunov exponent
can be calculated also as
\begin{equation}
\lambda=\left< \log|F'(x)| \right> = \left< \log|f'(x)| \right> \equiv
\int P_{st}(x) \log|f'(x)| dx
\label{lyapunov}
\end{equation}
Here we see clearly the two contributions to the Lyapunov
exponent: although the derivative $f'(x)$ does not change when
including noise in the trajectory, the stationary probability does
change (see Fig.4), thus producing the observed change in the
Lyapunov exponents. Synchronization, then, can be a general
feature in maps, such as (\ref{eq:3}), which have a large region
in which the derivative $|f'(x)|$ is smaller than one. Noise will
be able to explore that region and yield, on the average, a
negative Lyapunov exponent. This is, basically, the argument
developed in \cite{LZ98}.

\begin{figure}[htb]
\epsfig{file=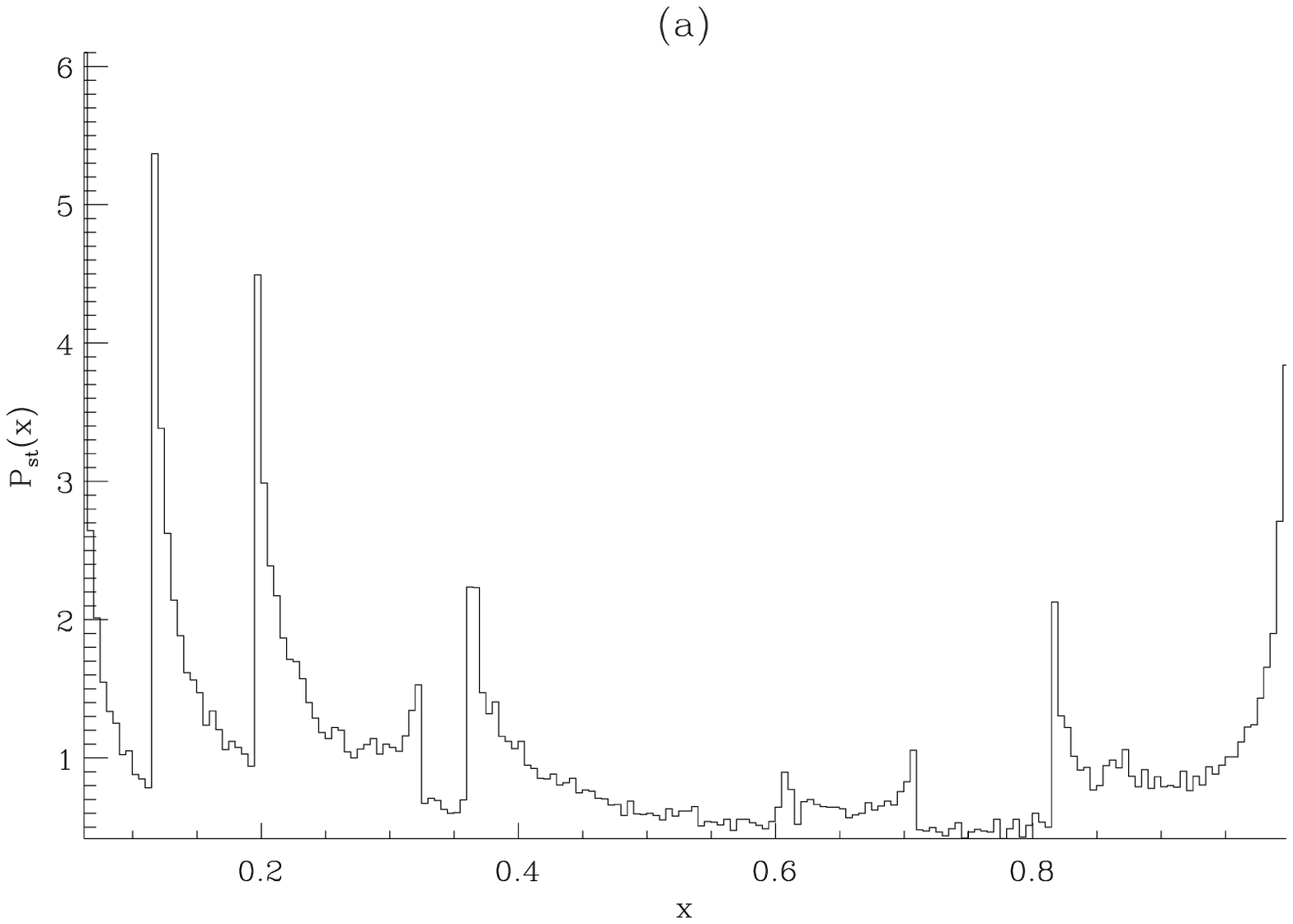,width=6.0cm}
\epsfig{file=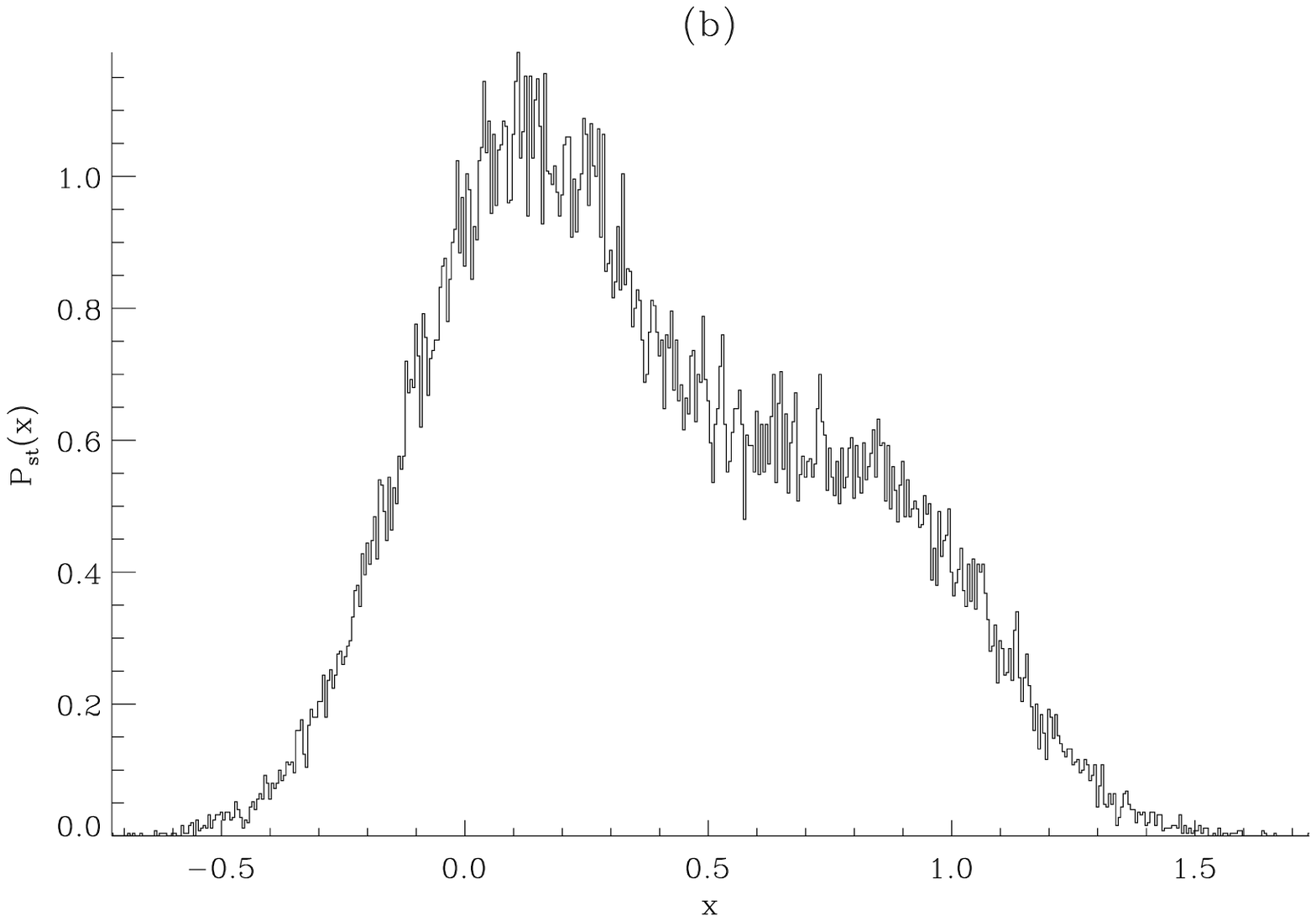,width=6.0cm}
\caption{
\label{fig4}
Plot of the stationary distribution for the map given by Eqs.(\ref{eq:2}) and
(\ref{eq:3}) with $\omega=0.3$
in the (a) deterministic case
$\epsilon=0$, and (b) the case with noise along the trajectory,
$\epsilon=0.2$.}
\end{figure}

In order to make some analytical calculation that can obtain in a
rigorous way the transition from a positive to negative Lyapunov
exponent, let us consider the map given by Eq. (\ref{eq:2}) and
\begin{equation}
       f(x) = \left\{
       \begin{array}{ll}
               a(1-\exp(1+x))   & \mbox{if $x < -1$}         \\
              -2-2x           & \mbox{if $x \in (-1,-.5)$}   \\
              2x              & \mbox{if $x \in (-.5,.5)$}   \\
               2-2x           & \mbox{if $x \in (.5,1)$}     \\
              a(-1+\exp(1-x))  & \mbox{if $x > 1$}
       \end{array}\right.
                  \label{twotents}
\end{equation}
\\
with $0<a<1$. This particular map, based in the tent map
\cite{tent}, has been chosen just for convenience. The following
arguments would apply to any other map that in the absence of
noise takes most frequently values in the region with the highest slopes, but
which visits regions of smaller slope when noise is introduced.
This is the case, for example, of the map (3). In the case of
(\ref{twotents}), the values given by the deterministic part of
the map, after one iteration from arbitrary initial conditions,
fall always in the interval $(-1,1)$. This is the region with the
highest slope $|F'|=2$. In the presence of noise the map can take
values outside this interval and, since the slopes encountered are
smaller, the Lyapunov exponent can only be reduced from the
deterministic value. To formally substantiate this point, it is
enough to recall the definition of Lyapunov exponent
(\ref{lyapunov2}): an upper bound for $|F'(x)|$ is $2$, so that a
bound for $\lambda$ is immediately obtained: $\lambda
\leq \ln 2$. Equality is obtained for zero noise.

The interesting point about the map (\ref{twotents}) and similar
ones is that one can demonstrate analytically that $\lambda$ can
be made negative. The intuitive idea is that it is enough to
decrease $a$ in order to give arbitrarily small values  to the
slopes encountered outside $(-1,1)$, a region accessible only
thanks to noise. To begin with, let us note that $|F'(x)|=2$ if $x
\in (-1,1)$, and $|F'(x)|< a$ if $|x|> 1$, so that an upper  bound
to (\ref{lyapunov2}) can be written as
\begin{eqnarray}
\nonumber
\lambda & \leq & \lim_{N\rightarrow\infty}  \left( \frac{N_I}{N} \ln 2
+ \frac{N_O}{N}\ln a \right)= p_I \ln 2 + p_O \ln a \\ & = & \ln 2 - p_O
\ln(2/a).
\label{bound}
\end{eqnarray}
$N_I/N$ and $N_O/N$ are the proportion of values of the map inside
$I=(0,1)$ and outside this interval, respectively, and we have
used that as $N\rightarrow\infty$ they converge to $p_I$ and
$p_O$, the invariant measure associated to $I$ and to the rest of
the real line, respectively ($p_I+p_O=1$).  A sufficient condition
for $x_{n+1}=f(x_n)+\epsilon \xi_n$ to fall outside $I$ is that
$|\xi_n|>2/\epsilon$. Thus, $p_O={\rm Probability}(|x_{n+1}|>1) >
{\rm Probability}(|\xi_{n}|>2/\epsilon) ={\rm
erfc}(\sqrt{2}/\epsilon)
\equiv T$, where we have used the Gaussian character of the noise.
In consequence, one finds from (\ref{bound})
\begin{equation}
\lambda \leq  \ln 2 - T \ln(2/a).
\label{rebound}
\end{equation}
The important point is that $T={\rm erfc}(\sqrt{2}/\epsilon)$ is
independent on the map parameters, in particular on $a$. Thus,
(\ref{rebound}) implies that by decreasing $a$ the value of
$\lambda$ can be made as low as desired. By increasing $\epsilon$
such that $T>\ln 2/\ln(2/a)$, $\lambda$ will be certainly
negative. Thus we have shown analytically that strong enough noise
will always make negative the Lyapunov exponent of the map
(\ref{twotents}) and, accordingly, it will induce yield
``noise-induced synchronization'' in that map.

\section{The Lorenz System}
\label{lorenz}
In this section we give yet another example of noise--induced
synchronization. We consider the well known Lorenz\cite{lorenz} model with
additional random terms of the form\cite{MB94}:

\begin{eqnarray}
\dot x & = & p(y-x) \nonumber \\
\dot y & = & -x z + r x -y +\epsilon \xi  \label{eq:lor}\\
\dot z & = & x y -b z \nonumber
\end{eqnarray}
\\
$\xi$ is white noise: a Gaussian random process of mean zero,
$\langle \xi(t)\rangle =0$ and delta--correlated, $\langle \xi(t)
\xi(t') \rangle = \delta (t-t')$. We have used $p=10$, $b=8/3$ and
$r=28$ which, in the deterministic case, $\epsilon=0$ are known to
lead to a chaotic behavior (the largest Lyapunov exponent is
$\lambda
\approx  0.9  >0$). As stated in the introduction, previous results seem to imply
that synchronization is
only observed for a noise with a non--zero mean. However, our results show
otherwise.

We have integrated numerically the above equations using the
stochastic Euler method \cite{RaulEuler}. Specifically, the evolution algorithm
reads:
\begin{eqnarray}
x(t+\Delta t) & = & x(t) + \Delta t \left[ p(y(t)-x(t))\right] \nonumber \\
y(t+\Delta t) & = & y(t) + \Delta t \left [-x(t) z(t) + r x(t) -y(t) \right] \label{eq:euler}\\
  &  + & \epsilon \sqrt{\Delta t} g(t) \nonumber \\
z(t+\Delta t) & = & z(t)+ \Delta t \left[ x(t) y(t) -b z(t)\right] \nonumber
\end{eqnarray}
\\The values of $g(t)$ are drawn at each time step from an
independent Gaussian distribution of zero mean and variance one
and they have been generated by a particularly efficient algorithm
using a numerical inversion technique\cite{TC93}. The time step
used is $\Delta t = 0.001$ and simulations range typically for a
total time of the order of $t=10^4$ (in the dimensionless units of
the Lorenz system of equations). The largest Lyapunov exponent has
been computed using a simultaneous integration of the linearized
equations\cite{PC89}. For the deterministic case, trajectories
starting with different initial conditions are completely
uncorrelated, see Fig. (5a). This is also the situation for small
values of $\epsilon$. However, when using a noise intensity
$\epsilon=40$ the noise is strong enough to induce synchronization
of the trajectories. Again, the presence of the noise terms forces
the largest Lyapunov exponent to become negative (for
$\epsilon=40$ it is $\lambda
\approx -0.2$).  As in the examples of the maps, after some
transient time, two different evolutions which have started in
completely different initial conditions synchronize towards the
same value of the three variables (see Fig. (5b) for the $z$
coordinate). Therefore, these results prove that synchronization
by common noise in the chaotic Lorenz system does occur for
sufficiently large noise intensity. This result contradicts
previous ones in the literature\cite{M96,SMP97}. The main
difference with these papers is in the intensity of the noise: it
has to be taken sufficiently large, as here, in order to observe
synchronization. Notice that although the noise intensity is
large, the basic structure of the ``butterfly" Lorenz attractor
remains present as shown in Fig. (6). Again, this result shows
that, although the noise intensity used could be considered large,
the synchronization is rather different from what would be
obtained from a trivial common synchronization of both systems to
the noise variable by neglecting the deterministic terms.

\section{Structural stability}
\label{structural}
An important issue concerns the structural stability of this
phenomenon, in particular how robust is noise synchronization to
small differences between the two systems one is trying to
synchronize. Whether or not the synchronization of two
trajectories of the same noisy Lorenz system (or of any other
chaotic system) observed here, equivalent to the synchronization
of two identical systems driven by a common noise, could be
observed in the laboratory, depends on whether the phenomenon is
robust when allowing the two Lorenz systems to be not exactly
equal (as they can not be in a real experiment). If one wants to
use this kind of stochastic synchronization in electronic emitters
and receivers (for instance, as a means of encryption) one should
be able to determine the allowed discrepancy between circuits
before the lack of synchronization becomes unacceptable.
Additional discussions on this issue may be found in
\cite{Pik92,MA98,yoshimura99}.

\begin{figure}[htb]
\epsfig{file=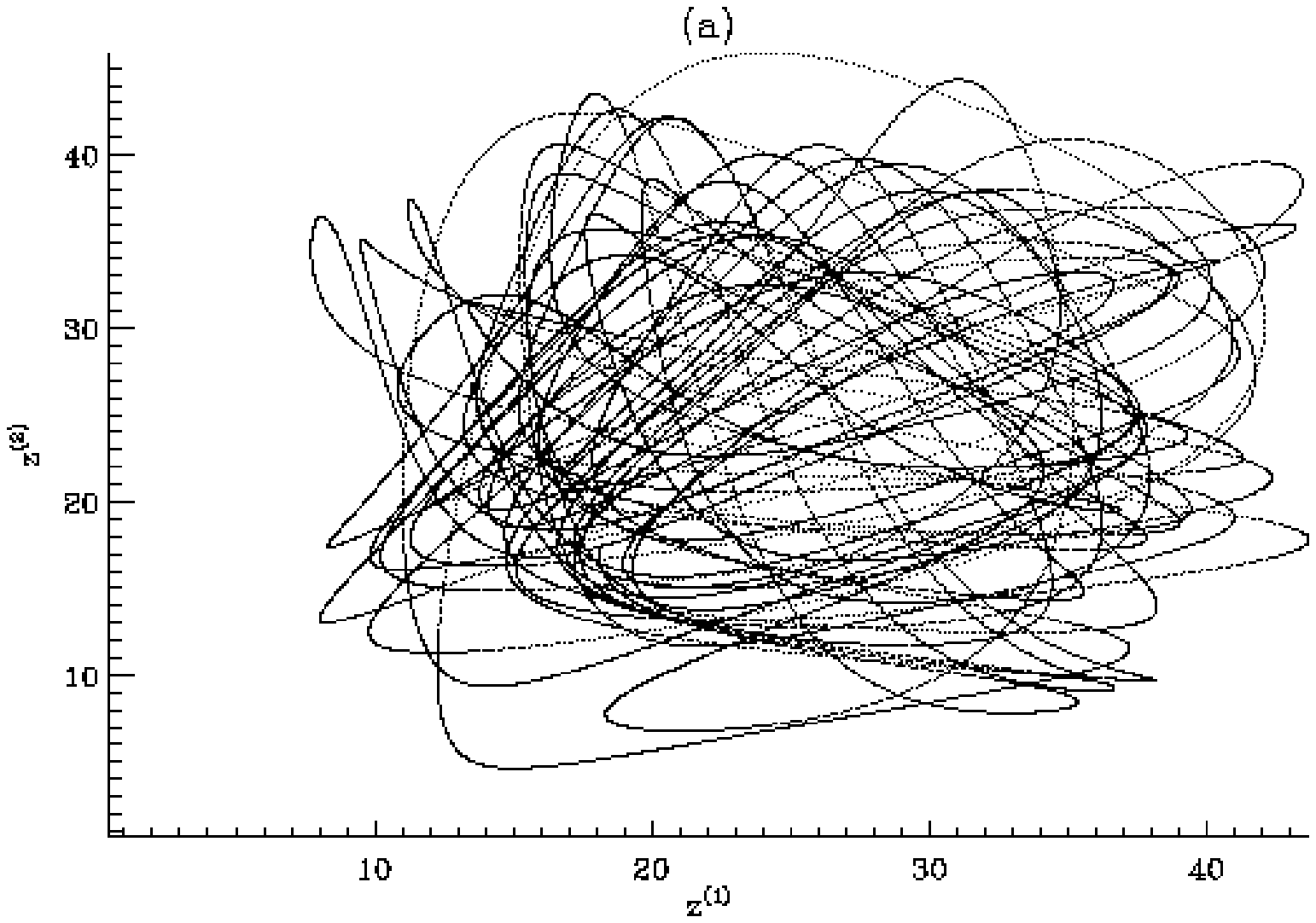,width=6.0cm}
\epsfig{file=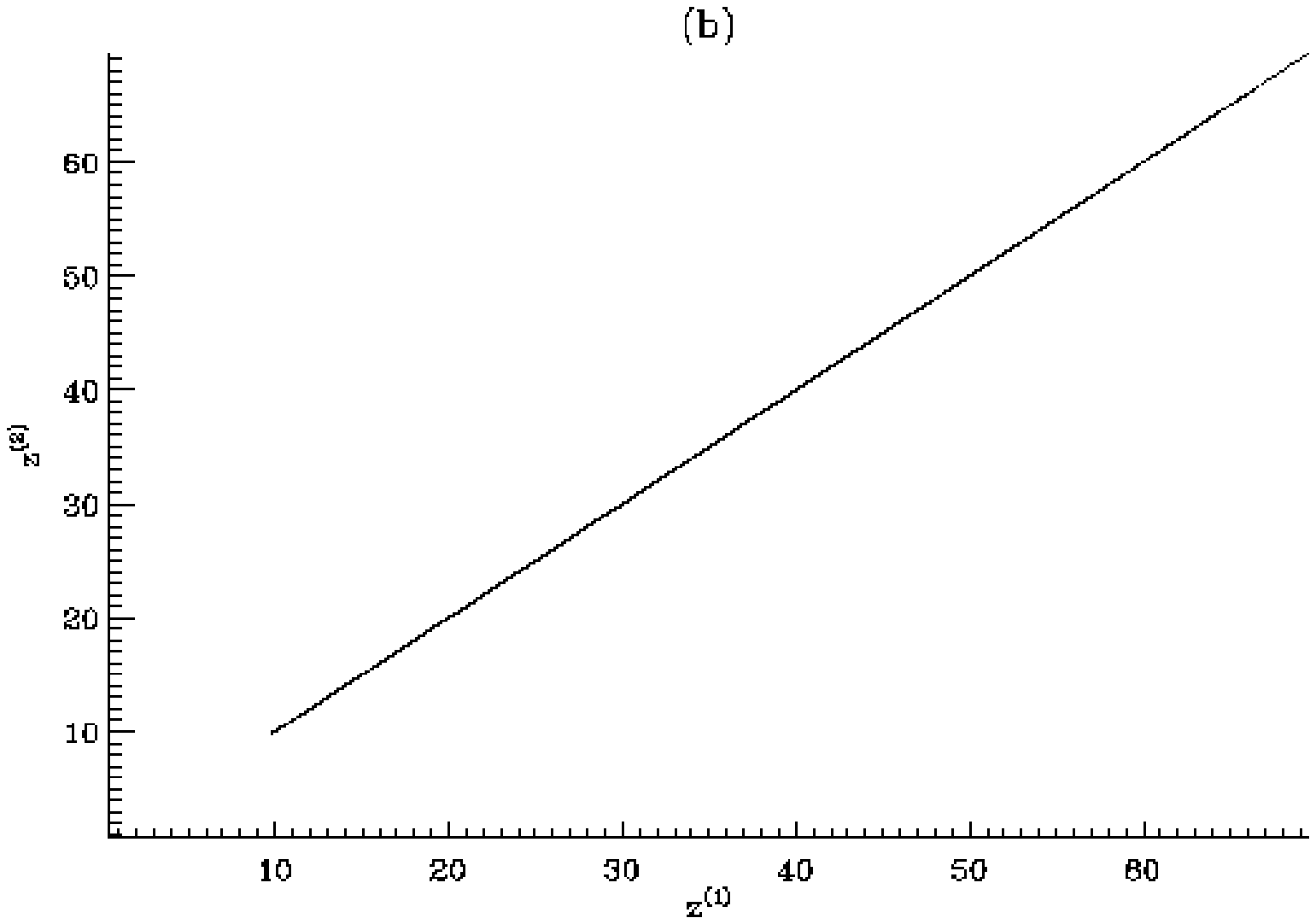,width=6.0cm}
\caption{ \label{fig5} Plot of two realizations $z^{(1)}$,
$z^{(2)}$ of the Lorenz system Eqs.(\ref{eq:lor}) with $p=10$,
$b=8/3$ and $r=28$. Each plotted realization starts from a
different initial condition and consists of an initial warming up
time of $t=12000$ (not shown in the figure) and runs for a time
$t=600$ in the dimensionless units of the Lorenz system of
equations. Panel (a) shows the deterministic case ($\epsilon=0$)
and panel (b) shows the results for $\epsilon=40$. Notice the
perfect synchronization in case (b).}
\end{figure}

\begin{figure}[ht]
\epsfig{file=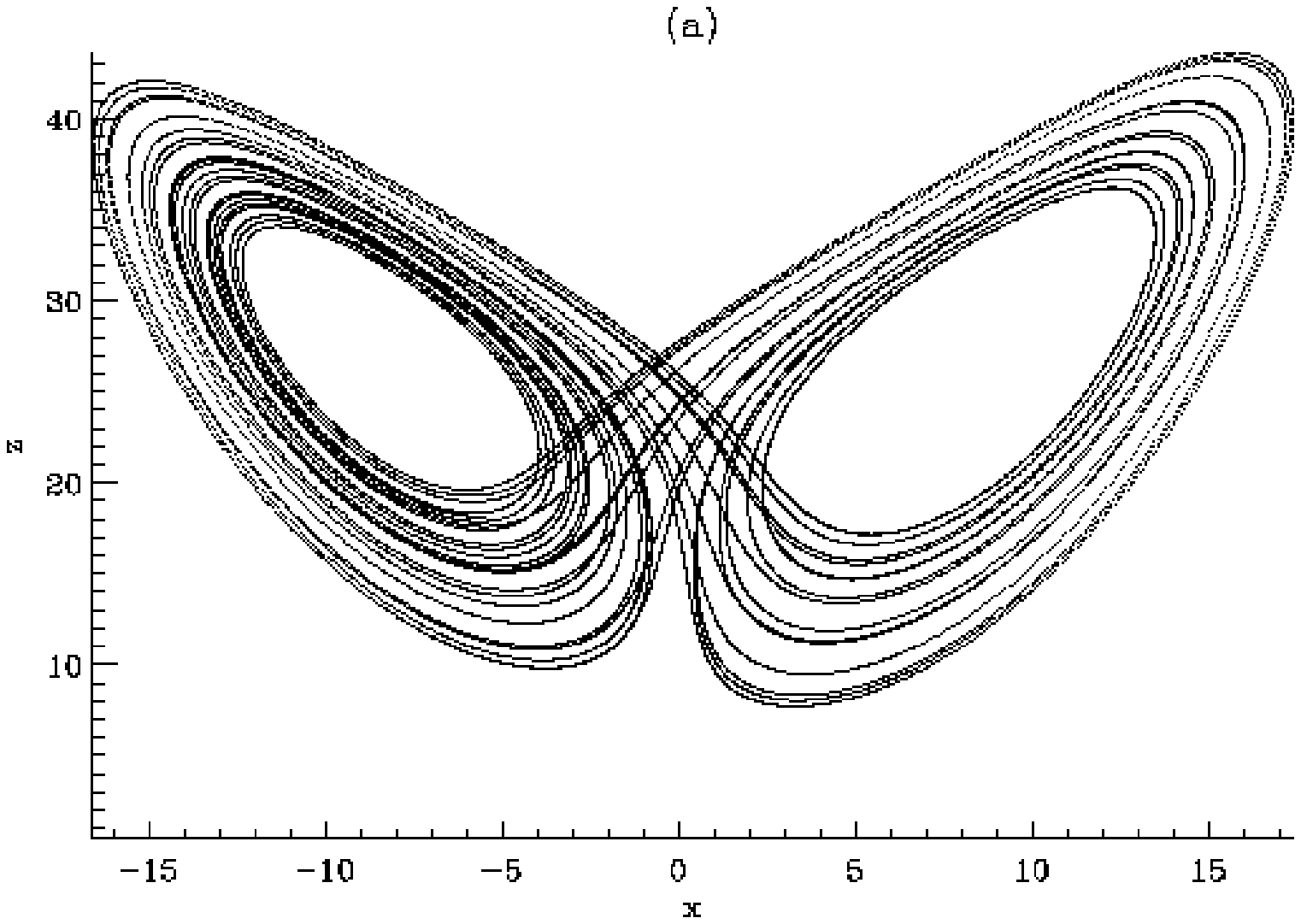,width=6.0cm}
\epsfig{file=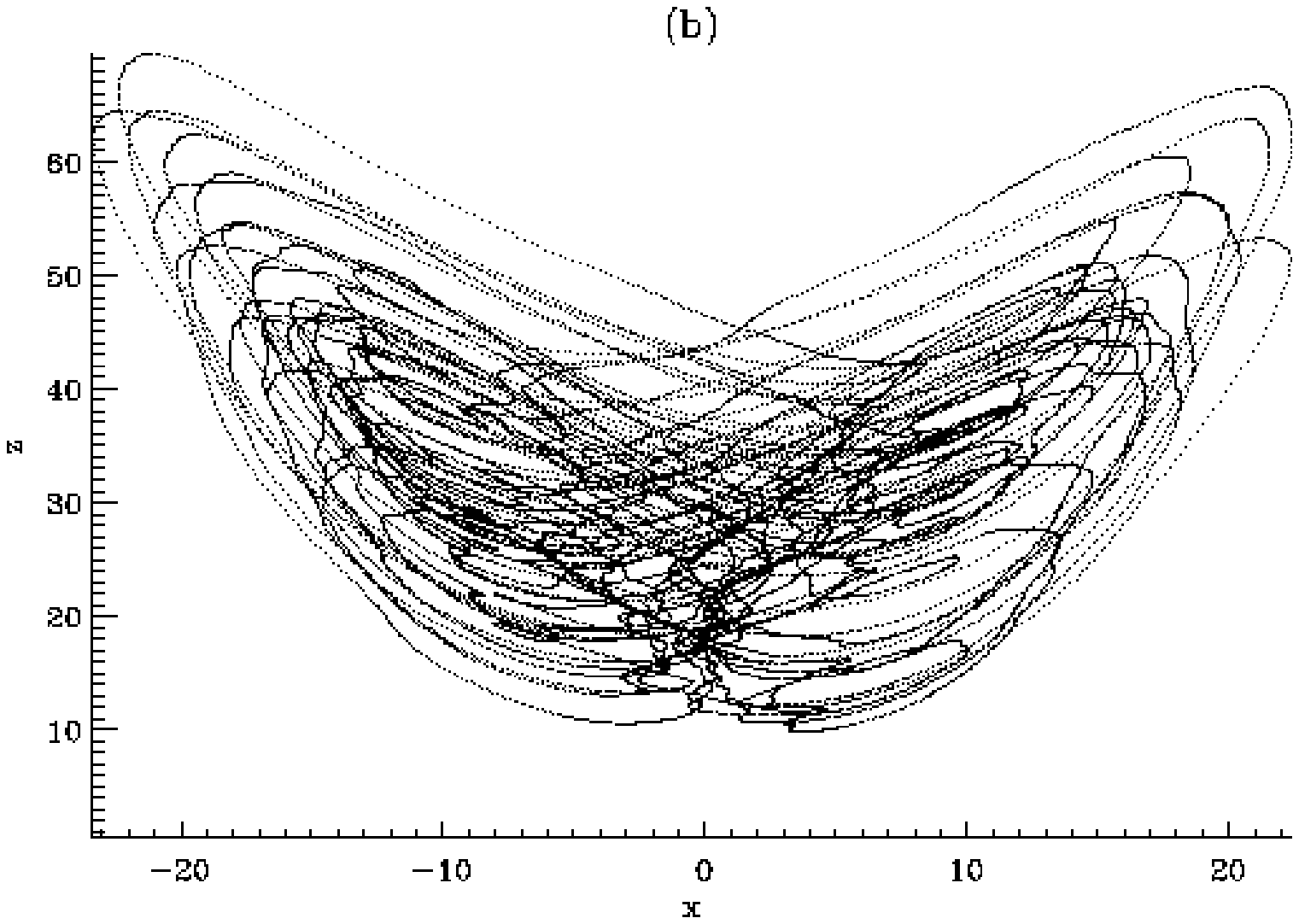,width=6.0cm}
\caption{ \label{fig6} ``Butterfly" attractor of the Lorenz system
in the cases (a) of no noise $\epsilon=0$, and (b) $\epsilon=40$
using the same time series as in figure 5. }
\end{figure}

We consider the following two maps forced by the same noise:
\begin{eqnarray}
x_{n+1}& = & f(x_n)+\xi_n
\label{first} \\
y_{n+1}& =& g(y_n)+\xi_n
\label{second}
\end{eqnarray}
 Linearizing in
the trajectory difference $u_n=y_n-x_n$, assumed to be small, we
obtain
\begin{equation}
u_{n+1}=g'(x_n) u_n + g(x_n)-f(x_n) \equiv g'(x_n) u_n +
\Delta(x_n)
\label{linearized}
\end{equation}
We have defined $\Delta(x) \equiv g(x)-f(x)$, and we are
interested in the situation in which the two systems are just
slightly different, for example, because of a small parameter
mismatch, so that $\Delta$ will be small in some sense specified
below.

Iteration of (\ref{linearized}) leads to the formal solution:
\begin{equation}
u_n=M(n-1,0) u_0 + \sum_{m=0}^{n-1} M(n-1,m+1)
\Delta(x_m)
\label{solution}
\end{equation}
We have defined $M(j,i)=\prod_{k=i}^j g'(x_k)$, and
$M(i-1,i)\equiv 1$. An upper bound on (\ref{solution}) can be
obtained:
\begin{equation}
\left|u_n\right|^2 \leq \left|M(n-1,0)\right|^2 \left|u_0\right|^2 +
 \sum_{m=0}^{n-1} \left|M(n-1,m+1)\right|^2 \left|\Delta(x_m)\right|^2
\label{bound2}
\end{equation}
The first term in the r.h.s. is what would be obtained for
identical dynamical systems. We know that $M(n-1,0)
\rightarrow e^{\lambda n}$ as $n\rightarrow \infty$, where
$\lambda$ is the largest Lyapunov exponent associated to
(\ref{second}). We are interested in the situation in which
$\lambda<0$, for which this term vanishes at long times.
Further analysis is done first for the case in which $\Delta(x)$
is a bounded function (or $x$ is a bounded trajectory with
$\Delta$ continuous). In this situation, there is a real number
$\mu$ such that $|\Delta(x_m)| < \mu$. We then get:
\begin{equation}
\left|u_n\right|^2 \leq \mu^2
 \sum_{m=0}^{n-1} \left|M(n-1,m+1)\right|^2
\label{bound3}
\end{equation}
an unequality valid for large $n$. Let us now define $K=\max_x |g(x)|$, the
maximum slope of the function $g(x)$. A trivial bound is now obtained as:
\begin{equation}
\left|u_n\right|^2 \leq \mu^2
\frac{1-K^{2n}}{1-K^2}
\label{bound4}
\end{equation}
This can be further improved in the case $K<1$, where we can write:
\begin{equation}
\left|u_n\right|^2 \leq \mu^2
\frac{1}{1-K^2}
\label{bound5}
\end{equation}
As a consequence, differences in the trajectories remain bounded
at all iteration steps $n$. Since, according to the definition
(\ref{lyapunov2}), $\ln K$ is also an upper bound for the Lyapunov
exponent for all values of $\epsilon$ and, in particular, for the
noiseless map, $\epsilon=0$, this simply tells us that if the
deterministic map is non--chaotic, then the addition of a common
noise to two imperfect but close replicas of the map will still
keep the trajectory difference within well defined bounds. The
situation of interest here, however, concerns the case in which a
negative Lyapunov exponent arises only as the influence of a
sufficiently large noise term, i.e. the deterministic map is
chaotic and $K>1$. In this case, the sum in Eq. (\ref{bound3})
contains products of slopes which are larger or smaller than $1$.
It is still true that the terms in the sum for large value of
$n-m$ can be approximated by $M(n-1,m+1)\approx {\rm
e}^{(n-m-1)\lambda}$ and, considering this relation to be valid
for all values of $n,m$, we would get:
\begin{equation}
 \sum_{m=0}^{n-1} \left|M(n-1,m+1)\right|^2
 \approx e^{2\lambda n}\sum_{m=0}^{n-1} e^{-2\lambda (m+1)} = \\
{ 1-e^{2\lambda n}  \over  1-e^{2\lambda} } \ .
\label{sumboundapprox}
\end{equation}
and, thus, at large $n$:
\begin{equation}
\left|u_n\right|^2 \lesssim \mu^2 (1-e^{2\lambda})^{-1}
\label{final}
\end{equation}
It can happen, however, that the product defining $M(n-1,m+1)$
contains  a large sequence of large slopes $g'(x_i)$. These terms
(statistically rare) will make the values of $|u_n|$ to violate
the above bound at sporadic times. Analysis of the statistics of
deviations from synchronization was carried out in \cite{Pik92}.
Although for $\lambda<0$ the most probable deviation is close to
zero, power-law distributions with long tails are found, and
indeed its characteristics are determined by the distribution of
slopes encountered by the system during finite amounts of time, or
finite-time Lyapunov exponents, as the arguments above suggest.
Therefore, we expect a dynamics dominated by relatively large
periods of time during which the difference between trajectories
remains bounded by a small quantity, but intermittently
interrupted by bursts of large excursions of the difference. This
is indeed observed in the numerical simulations of the maps
defined above. This general picture is still valid even if
$|\Delta(x)|$ is not explicitly bounded.

We have performed a more quantitative study for the case in which
two noisy Lorenz systems with different sets of parameters,
namely:
\begin{eqnarray}
\dot x_1 & = & p_1(y_1-x_1) \nonumber \\
\dot y_1 & = & -x_1 z_1 + r_1 x_1 -y_1 +\epsilon \xi  \label{eq:lor1}\\
\dot z_1 & = & x_1 y_1 -b_1 z_1 \nonumber
\end{eqnarray}
\\and
\begin{eqnarray}
\dot x_2 & = & p_2(y_2-x_2) \nonumber \\
\dot y_2 & = & -x_2 z_2 + r_2 x_2 -y_2 +\epsilon \xi  \label{eq:lor2}\\
\dot z_2 & = & x_2 y_2 -b_2 z_2 \nonumber
\end{eqnarray}
\\
are forced by the same noise $\xi(t)$. In order to discern the
effect of each parameter separately, we have varied independently
each one of the three parameters, $(p,b,r)$, while keeping
constant the other two. The results are plotted in Fig. 7. In this
figure we plot the percentage of time in which the two Lorenz
systems are still synchronized with a tolerance of 10\%. This
means that trajectories are considered synchronized if the
relative difference in the $z$ variable is less than 10\%.
According to the general discussion for maps, we expect departures
from approximate synchronization from time to time. They are in
fact observed, but from Fig. 7 we conclude that small variations
(of the order of 1\%) still yield a synchronization time of more
than 85\%. In Fig. 8 we show that the loss of synchronization
between the two systems appears in the form of bursts of spikes
whose amplitude is only limited by the size of the attractor in
the phase space. Moreover, it can be clearly seen in the same
figure that large (but infrequent) spike amplitudes appear for
arbitrarily small mismatch.

\begin{figure}[ht]
\epsfig{file=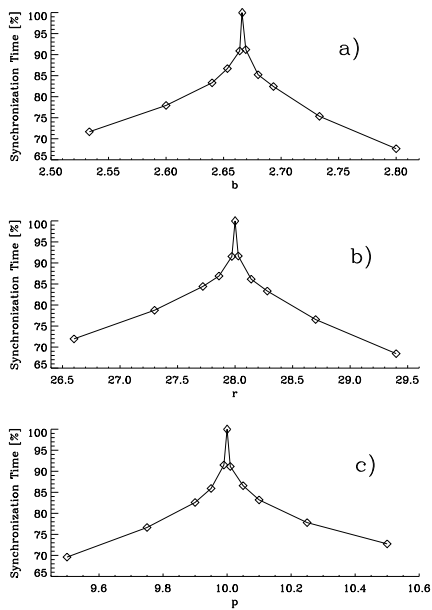,width=6.0cm}
\caption{ \label{fig7} Percentage of time that two slightly
dissimilar Lorenz systems, subjected to the same noise, remain
synchronized (up to a 10\% discrepancy in the $z$ variable). For
one of the two systems we fix $p_1=10$, $b_1=8/3$ and $r_1=28$ while for
the other we vary systematically one of the parameters keeping the
other two constant: in panel (a) the parameter $b_2$ varies , in
panel (b) the parameter $r_2$ varies and in panel (c) the parameter
$p_2$ varies. Notice that the percentage of synchronization time is
still higher than 85\% if the relative difference between the
parameters is less than 1\%.}
\end{figure}

In the realm of synchronization of chaotic oscillators, two
different types of analogous intermittent behavior have been
associated also to the fluctuating character of the finite-time
conditional Lyapunov exponents as above. One is on-off
intermitency \cite{ONOFF} where the synchronization manifold is
sligthly unstable on average but the finite time Lyapunov exponent
is negative during relatively long periods of time. In the other
one, named bubbling \cite{BUBBLING}, the synchronization is stable
on average but the local conditional Lyapunov exponent becomes
occasionally positive. While in the former case bursting always
occurs due to the necessarily imperfect initial synchronization,
in the latter it is strictly a consequence of the mismatch of the
entraining systems. In this sense, the behavior reported in the
preceding paragraph should be considered as a manifestation of
bubbling in synchronization by common noise.

\begin{figure}[ht]
\epsfig{file=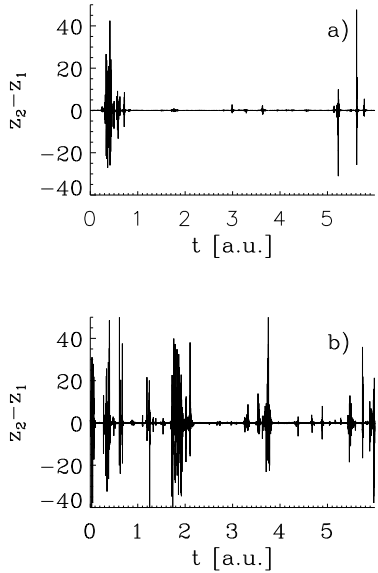,width=6.0cm}
\caption{\label{fig8} Time evolution of the difference between two
trajectories $z_1$ and $z_2$ corresponding to two Lorenz systems
driven by the same noise but with a small mismatch in the $r$
parameter: panel (a) $r_1=28$ and $r_2=0.99999 \, \times \, r_1$
and in panel (b) $r_2=0.999 \, \times \, r_1$. Notice that
although the synchronization time diminishes with increasing
parameter mismatch, the maximum absolute difference between the
two variables does not depend on the mismatch.}
\end{figure}

\section{Conclusions and open questions}
\label{conclusions}

In this paper we have addressed the issue of synchronization of
chaotic systems by the addition of common random noises. We have
considered three explicit examples: two 1-d maps and the Lorenz
system under the addition of zero--mean, Gaussian, white noise.
While the map examples confirm previous results in similar maps,
and we have obtained with them analytical confirmation of the
phenomenon, the synchronization observed in the Lorenz system
contradicts some previous results in the literature. The reason is
that previous works considered noise intensities smaller than the
ones we found necessary for noise-synchronization in this system.
Finally, we have analyzed the structural stability of the observed
synchronization. In the Lorenz system, synchronization times
larger than 85\% (within an accuracy of 10\%) can still be
achieved if the parameters of the system are allowed to change in
less than 1\%.

It is important to point out that noise-induced synchronization
between identical systems subjected to a common noise is
equivalent to noise induced order, in the sense that the Lyapunov
exponent defined in (4) becomes negative in a single system
subjected to noise. One can ask whether the state with negative
Lyapunov exponent induced by noise may be still be called
`chaotic' or not. This is just a matter of definition: if one
defines chaos as exponential sensibility to initial conditions,
and one considers this {\sl for a fixed noise realization}, then
the definition of Lyapunov exponent implies that trajectories are
not longer chaotic in this sense. But one can also consider the
extended dynamical system containing the forced one {\sl and} the
noise generator (for example, in numerical computations, it would
be the computer random number generator algorithm). For this {\sl
extended system} there is strong sensibility to initial conditions
in the sense that small differences in noise generator seed leads
to exponential divergence of trajectories. In fact, this
divergence is at a rate given by the Lyapunov exponent of the
noise generator, which approaches infinity for a true Gaussian
white process. Trajectories in the noise-synchronized state are in
fact more irregular than in the absence of noise, and attempts to
calculate the Lyapunov exponent just from the observation of the
time series will lead to a positive and very large value, since it
is the {\sl extended} dynamical system the one which is observed
when analyzing the time series \cite{vulpibook} (typically such
attempts will fail because the high dimensionality of good noise
generators, ideally infinity, would put them out of the reach of
standard algorithms for Lyapunov exponent calculations). Again,
whether or not to call such irregular trajectories with just
partial sensibility to initial conditions `chaotic' is just a
matter of definition. More detailed discussion along these lines
can be found in \cite{vulpi95}.

There remain still many open questions in this field. They
involve the development of a general theory, probably based in the
invariant measure, that could give us a general criterion to
determine the range of parameters (including noise levels) for
which the Lyapunov exponent becomes negative, thus allowing
synchronization. In this work and similar ones, the word
synchronization is used in a very restricted sense, namely: the
coincidence of asymptotic trajectories. This contrasts with the
case of interacting periodic oscillations where a more general
theory of synchronization exists to explain the phenomenon of non
trivial phase locking between oscillators that individually
display very different dynamics. Indications of the existence of
analogue non trivial phase locking have been reported for chaotic
attractors\cite{RPK96}. There a ``phase" with a chaotic trajectory
defined in terms of a Hilbert transform is shown to be
synchronizable by external perturbations in a similar way as it
happens with periodic oscillators. Whether or not this kind of
generalized synchronization can be induced by noise is, however, a
completely open question. Last, but not least, it would be also
interesting to explore whether analogs of the recently reported
synchronization of spatio-temporal chaos\cite{AHMSM97,G99} may be
induced by noise.

{\bf Acknowledgements} We thank Changsong Zhou for useful
comments. Financial support from DGESIC (Spain), projects
PB97-0141-C02-01 and BFM2000-1108, is acknowledged.

\end{twocolumns}

\end{document}